\documentclass[aps,prl,floatfix,twocolumn,superscriptaddress]{revtex4-2}
\usepackage[normalem]{ulem}
\usepackage{bm}
\usepackage{color}

\usepackage{graphicx,
            caption,
            subcaption,
            amsmath,
            braket,
            units,
            ragged2e,
            xcolor,
            xspace,
            nicefrac,
            lipsum,
            nameref,
            textcomp,
            urwchancal,
            upgreek,
            isotope,
            hyperref,
            cleveref,
            siunitx,
            placeins, 
            scrextend, 
            url}

\DeclareCaptionLabelSeparator{dot}{. }
\makeatletter
\def\justified{
	\let\\\@normalcr
	\@rightskip\z@skip \rightskip\@rightskip
	\leftskip\z@skip
	\parindent 0em\relax
	\setlength{\parfillskip}{0pt plus 1fil}}
\DeclareCaptionJustification{justified}{\justified}
\hypersetup{colorlinks=true, linkcolor=blue,citecolor=blue,urlcolor=blue}
\def\unit #1 #2 {\SI{#1}{#2}\xspace}
\def\range #1 #2 #3 {\SIrange{#1}{#2}{#3}\xspace}
\DeclareSIUnit\gauss{G}

\newcommand{\myref}[2][]{Fig.~\hyperref[#2]{\ref*{#2}#1}}
\newcommand{\Myref}[2][]{Figure~\hyperref[#2]{\ref*{#2}#1}}
\newcommand{\Mytabref}[2][]{Table~\hyperref[#2]{\ref*{#2}#1}}

\begin{document}

\title{Can angular oscillations probe superfluidity in dipolar supersolids?}

 \author{Matthew A. Norcia}
 \affiliation{
     Institut f\"{u}r Quantenoptik und Quanteninformation, \"Osterreichische Akademie der Wissenschaften, Innsbruck, Austria
 }

\author{Elena Poli}
 \affiliation{
     Institut f\"{u}r Experimentalphysik, Universit\"{a}t Innsbruck, Austria
 }
 
 \author{Claudia Politi}
 \affiliation{
     Institut f\"{u}r Quantenoptik und Quanteninformation, \"Osterreichische Akademie der Wissenschaften, Innsbruck, Austria
 }
 \affiliation{
     Institut f\"{u}r Experimentalphysik, Universit\"{a}t Innsbruck, Austria
 }

 \author{Lauritz Klaus}
 \affiliation{
     Institut f\"{u}r Quantenoptik und Quanteninformation, \"Osterreichische Akademie der Wissenschaften, Innsbruck, Austria
 }
 \affiliation{
     Institut f\"{u}r Experimentalphysik, Universit\"{a}t Innsbruck, Austria
 }

\author{Thomas Bland}
 \affiliation{
     Institut f\"{u}r Quantenoptik und Quanteninformation, \"Osterreichische Akademie der Wissenschaften, Innsbruck, Austria
 }
 \affiliation{
     Institut f\"{u}r Experimentalphysik, Universit\"{a}t Innsbruck, Austria
 }
 
 \author{Manfred J. Mark}
 \affiliation{
     Institut f\"{u}r Quantenoptik und Quanteninformation, \"Osterreichische Akademie der Wissenschaften, Innsbruck, Austria
 }
 \affiliation{
     Institut f\"{u}r Experimentalphysik, Universit\"{a}t Innsbruck, Austria
 }

 \author{Luis Santos}
 \affiliation{
     Institut f\"{u}r Theoretische Physik, Leibniz Universit\"{a}t Hannover, Germany
 }
 
\author{Russell N. Bisset}
 \affiliation{
     Institut f\"{u}r Experimentalphysik, Universit\"{a}t Innsbruck, Austria
 }
 
 \author{Francesca Ferlaino}
 \thanks{Correspondence should be addressed to \mbox{\url{Francesca.Ferlaino@uibk.ac.at}}}
 \affiliation{
     Institut f\"{u}r Quantenoptik und Quanteninformation, \"Osterreichische Akademie der Wissenschaften, Innsbruck, Austria
 }
 \affiliation{
     Institut f\"{u}r Experimentalphysik, Universit\"{a}t Innsbruck, Austria
 }

\begin{abstract}
    Angular oscillations can provide a useful probe of the superfluid properties of a system. Such measurements have recently been applied to dipolar supersolids, which exhibit both density modulation and phase coherence, and for which robust probes of superfluidity are particularly interesting.  So far, these investigations have been confined to linear droplet arrays.  Here, we explore angular oscillations in systems with 2D structure, which in principle have greater sensitivity to superfluidity.  Surprisingly, in both experiment and simulation, we find that the frequency of angular oscillations remains nearly unchanged  even when the superfluidity of the system is altered dramatically.  This indicates that angular oscillation measurements do not always provide a robust experimental probe of superfluidity with typical experimental protocols.  
\end{abstract}

\maketitle

Some of the most distinctive manifestations of superfluidity in ultracold quantum gases relate to their behavior under rotation.  These include the presence of quantized vortices \cite{matthews1999vortices, Madison2000vortex, zwierlein2005vortices} and persistent currents in ring-traps \cite{Ramanathan2011}, as well as shape-preserving angular oscillations associated with a ``scissors" mode \cite{guery1999scissors}.  Measurements of the scissors mode frequency have long been used to illuminate the superfluid properties of a variety of systems \cite{bohle1984new, marago2000observation, Wright2007, Bijnen2010collective, Ferrier2018}. With the recent advent of dipolar supersolids \cite{Boninsegni:2012, Lu:2015, Biallie:2018, Roccuzzo:2019, Tanzi:2019, Bottcher2019, Chomaz:2019} --- states that possess both the global phase coherence of a superfluid and the spatial density modulation of a solid --- the scissors mode provides a tempting way to quantify changes in superfluidity across the superfluid-supersolid transition \cite{Roccuzzo:2020, tanzi2021evidence}.

\begin{figure}[!htb]
\includegraphics[width=3.38 in, ]{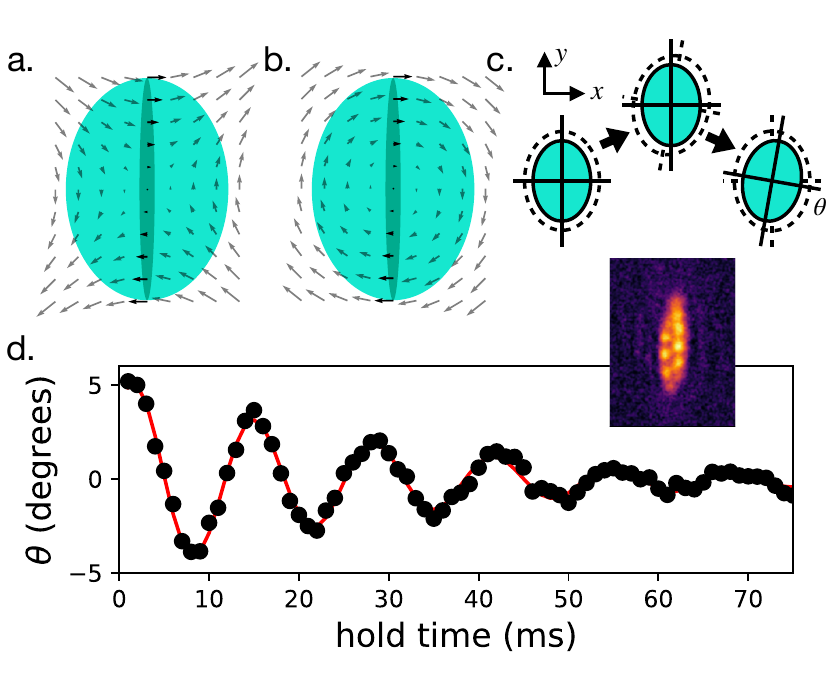}
\caption{Characteristic velocity profiles for irrotational flow (\textbf{a.}) and rigid-body rotation (\textbf{b.}). A wide atomic state (light turquoise oval) samples a region of space where the two differ significantly, while a highly elongated state (dark turquoise oval) samples a region where the two patterns are nearly indistinguishable. \textbf{c.} We excite oscillations in the angle $\theta$ of our atomic gas by rapidly rotating the anisotropic trap (dashed oval), then returning it to its original orientation and observing the subsequent dynamics.  \textbf{d.} Typical example of experimental angular oscillation for the zig-zag modulated state shown on the right (image averaged over 9 iterations).  In this case, the errors from the fit to the state angle are smaller than the markers.  The red line is a damped sinusoidal fit used to extract the angular oscillation frequency $f_{\rm{osc}}$. }
\label{fig:overview}
\end{figure}

The goal of these angular oscillation measurements is to infer the flow patterns allowed for a given fluid.  A superfluid is constrained by the single-valued nature of its wavefunction to irrotational flow (IF), while a non-superfluid system faces no such constraint, and in certain situations may be expected to undergo rigid-body rotation (RBR).  Prototypical velocity fields for angular oscillations under IF ($\vec{v} \propto \nabla x y$) and RBR ($\vec{v} \propto r \hat{\theta}$) are depicted in Fig.~\ref{fig:overview}a,b, respectively.  
The velocity field associated with angular rotation is related to the moment of inertia of the system, and thus the frequency of angular oscillations.

The ability to distinguish between RBR and IF (and thus, in principle, between a classical and superfluid system) depends critically on the geometry of the system, and is sensitive only to the character of the flow pattern where the atomic density is appreciable.  As illustrated in Fig.~\ref{fig:overview}a,b), highly elongated states sample only the region along the weak axis of the trap (near $x=0$) where IF and RBR are identical for small rotations (dark turquoise regions), while rounder states (light turquoise regions) sample regions of space where the flow patterns differ significantly, and are thus far more sensitive to the irrotational constraint. 

The interplay between the change in superfluid character and in geometrical structure is particularly subtle for dipolar gases. While pioneering theoretical works in extended systems predict a change in the moment of inertia across the superfluid-supersolid-solid phase transition, due purely to the reduction of the superfluid fraction \cite{Leggett:1970}, dipolar gases additionally undergo a structural change when driven from an unmodulated condensate to a supersolid state, which also affects the system’s rotational behavior. Recent works have focused on systems that form a short linear chain of about two ``droplets" \footnote{Here, the word {``droplet"} refers to a high-density region, which is not necessarily self-bound.} 
in the supersolid regime \cite{Roccuzzo:2020, tanzi2021evidence}.  In these systems, the transition from an unmodulated BEC to a supersolid is accompanied by a dramatic narrowing of the atomic density distribution, which both leads to a change in the angular oscillation frequency and a reduced sensitivity to superfluidity in the modulated regime.  Further, because the motion induced by rotation in a linear system is perpendicular to the interdroplet axis, the scissors mode frequency may not be sensitive to the superfluid connection along the interdroplet axis, but rather to the low-density superfluid ``halo" that surrounds the droplets.  
These geometrical considerations appear more favorable for supersolids with two-dimensional structure.  In this case, a  relatively round aspect ratio is maintained across the transition, isolating the effects of superfluidity from geometry \cite{supmat}.  \\

In the present work, we study the angular response of recently demonstrated supersolids with two-dimensional structure \cite{norcia2021two, bland2021two}.  
With this more favorable geometry, one may expect to observe unambiguous effects of superfluidity on the frequency of oscillations following a sudden angular rotation.    However, we do not.  Surprisingly, we find that unmodulated fully superfluid condensates, supersolids, and isolated-droplet crystals all oscillate at a frequency very close to that predicted for a fully superfluid state.  This observation points to the elastic nature of the modulated phases of dipolar gases.  We extensively investigate the system behavior as a function of geometry and interaction parameters, revealing a unique multi-mode response of the dipolar supersolid.

Experimentally, we use a dipolar quantum gas of $^{164}$Dy atoms (up to approximately $5 \times 10^4$ condensed atoms), confined within an optical dipole trap (ODT) of tunable geometry, formed at the intersection of three laser beams \cite{Trautmann2018, norcia2021two, bland2021two}.  The trap geometry and particle number at the end of the evaporative cooling sequence determine the character of the modulated ground state, which can form linear, zigzag, or triangular lattice configurations \cite{poli2021maintaining}.  By varying the applied magnetic field in the vicinity of Feshbach resonances near 18-23~G, we can access scattering lengths that correspond to either unmodulated BECs or modulated states.  In past works, we have demonstrated that modulated states created at the corresponding field have global phase coherence \cite{norcia2021two, bland2021two}.  In this work, we expect the same to be true, but refer to these experimental states simply as modulated, as we do not repeat the characterization for every trap condition used.  We excite angular oscillations by using the well-established protocol of applying a sudden small rotation of the trap, by varying the relative powers in the ODT beams for 6~ms before returning them to their original values (Fig.~\ref{fig:overview}c). 
Using our high-resolution imaging \cite{Sohmen2021}, we observe the in-trap density profile at a variable time from the excitation, and extract the angle of the major and minor axes using a two-dimensional Gaussian fit to the state \cite{supmat}.  

\begin{figure}[!htb]
\includegraphics[width=3.38 in, ]{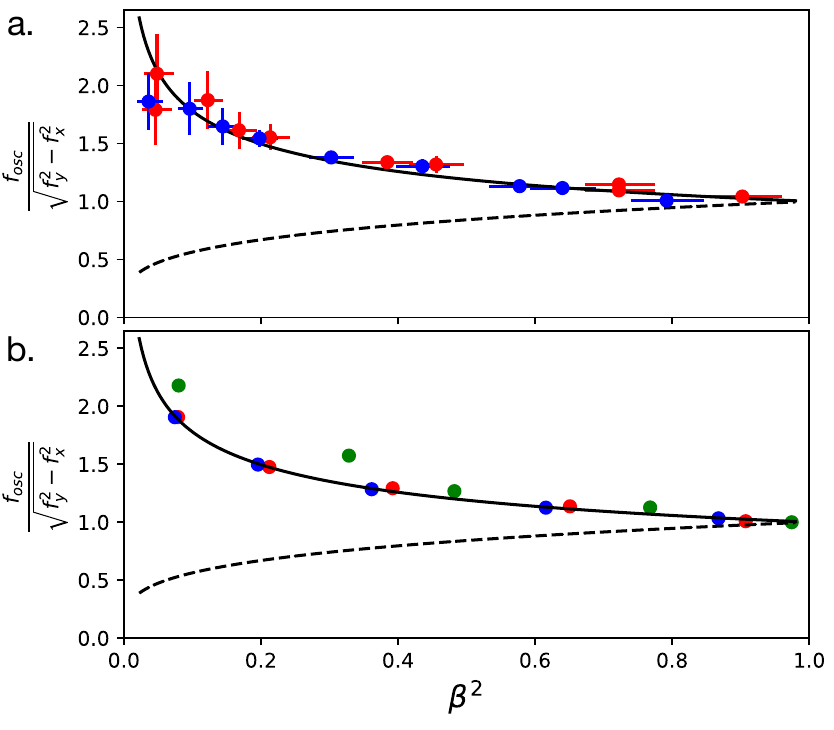}
\caption{Normalized oscillation frequencies $f_{\rm{osc}}$ from experiment (a) and simulation (b).  Blue points represent unmodulated BECs, red points represent modulated states (expt.) and supersolid states (sim.), and green points represent independent droplet arrays.  Solid lines are predictions for irrotational flow $f_{\rm{irr}}$.  Dashed lines are predictions for rigid body rotation $f_{\rm{rig}}$.}
\label{fig:vsbeta}
\end{figure}

A typical angular oscillation is shown in Fig.~\ref{fig:overview}d, for a ``zig-zag" modulated state \cite{norcia2021two}.  From such oscillation traces, we extract the dominant oscillatory frequency $f_{\rm{osc}}$ using a fit to an exponentially damped sinusoid.  Typically, the statistical error on our measurements of $f_{\rm{osc}}$ is on the sub-Hertz level, better than our knowledge of the trap frequencies, due to drifts between calibrations.  We perform such measurements for trap geometries ranging from an elongated cigar-shape to pancake-shaped, and for different scattering lengths, as summarized in Fig.~\ref{fig:vsbeta}a.

Within a single-mode approximation, the angular oscillation frequency $f_{\rm{osc}}$ can be predicted using either a sum-rule based approach \cite{pitaevskii2016bose, Roccuzzo:2020}, or considerations based on hydrodynamic flow \cite{guery1999scissors}.  For RBR, the angular oscillation frequency is given by  $f_{\rm{rig}} = \sqrt{(f_y^2 - f_x^2)\beta}$, whereas for IF, the predicted value is $f_{\rm{irr}} = \sqrt{(f_y^2 - f_x^2)/\beta}$ \cite{Roccuzzo:2020, tanzi2021evidence}.  Here, $f_{x,y}$ are the trap frequencies along directions $x$ and $y$. $\beta = \langle x^2 - y^2 \rangle / \langle x^2 + y^2\rangle $ is a geometrical factor that quantifies the degree of elongation of the atomic cloud. As shown in Fig.~\ref{fig:vsbeta}, $f_{\rm{rig}}$ and $f_{\rm{rig}}$ are more distinct for larger values of $\beta$.  
Remarkably, independent of trap geometry or the presence of modulation, we observe $f_{\rm{osc}}$ close to the IF prediction, and far from the RBR prediction when the two predictions differ appreciably.  

\begin{figure}[!htb]
\includegraphics[width=3.38 in, ]{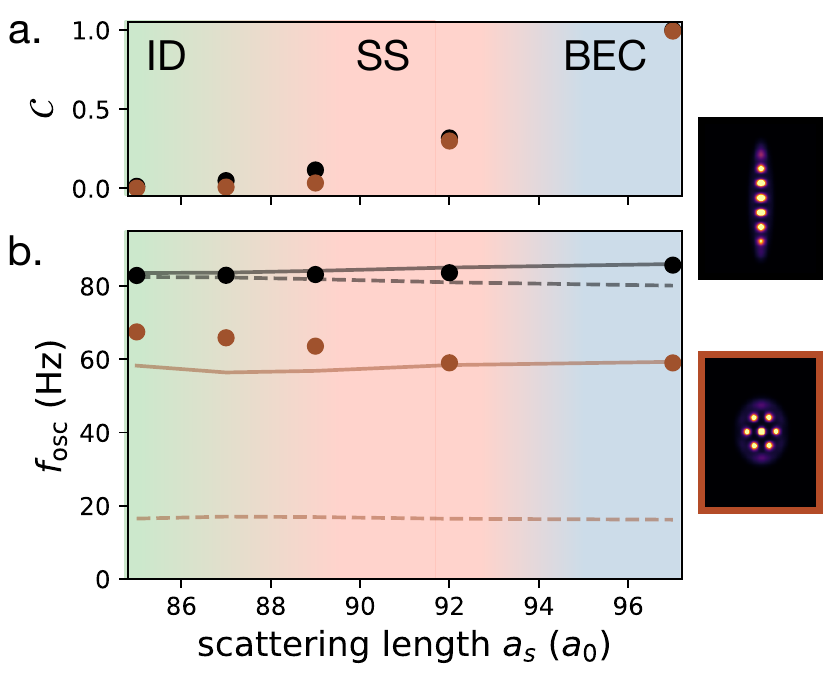}
\caption{Impact of scattering length on simulated scissors mode frequencies.  \textbf{a.} Interdroplet connection $\mathcal{C}$ (defined in text) versus scattering length for different trap geometries.  The calculated ground state in each trap is shown on the right, with corresponding border colors. \textbf{b.} Scissors mode frequency versus scattering length.  
Solid lines are predictions for irrotational flow $f_{\rm{irr}}$.  Dashed lines are predictions for rigid body rotation $f_{\rm{rig}}$.  $\beta$ ranges from 0.93 to 0.99, and 0.27 to 0.31 in the linear and hexagonal cases, respectively.  
}
\label{fig:vs_as}
\end{figure}
To gain a deeper understanding of this surprising observation, we theoretically study the oscillation dynamics using a real-time simulation of the extended Gross-Pitaevskii equation (eGPE) \cite{FerrierBarbut:2016, Chomaz:2016, Wachtler2016}.  To compare to the experimental observations of Fig.~\ref{fig:vsbeta}a, we first calculate the ground state for a given trap, scattering length and atom number.  We then apply a 0.5~degree rotation of the trap for 6~ms (we have confirmed that the character and frequency of the response do not change for much larger excitations), and then let the state evolve for 50~ms.  We then perform the same fitting procedure as used in the experiment to extract $f_{\rm{osc}}$.  For the simulation, we calculate $\beta$ directly for the ground state (we confirm that the exact value of $\beta$ agrees with that extracted from a Gaussian fit at the 5\% level).  For simulations performed on states ranging from the unmodulated BEC to supersolid to independent droplet regimes, with vanishing superfluid connection between droplets, we again find that $f_{\rm{osc}}$ is always very close to $f_{\rm{irr}}$, in very good agreement with the experimental data. For isolated droplet states in particular, $f_{\rm{osc}}$ can actually be even higher than the expected value for irrotational flow, indicating that the oscillation frequency is not necessarily in between the irrotational and rigid body values.

To further illuminate the dependence $f_{\rm{osc}}$ on superfluidity, we analyze the results of the simulation as a function of the s-wave scattering length $a_s$ (Fig.~\ref{fig:vs_as}).  Scattering lengths of 85~$a_0$ yield arrays of (nearly) independent droplets, while $a_s$ = 97~$a_0$ produces an unmodulated BEC.  In-between, we find supersolid states, with low-density connections between droplets.  Inspired by the formulation of Leggett \cite{Leggett:1970}, we quantify the degree of inter-droplet density connection as $\mathcal{C} = [\int dx /\rho(x)]^{-1}$, where $\rho(x)$ is the projected atomic density, evaluated over the inter-droplet connection (Fig.~\ref{fig:vs_as}a) \footnote{We note that the geometry of our system is very different from that considered in Ref.~\cite{Leggett:1970}, and so we do not expect this quantity to have a direct connection to the non-classical moment of inertia of the system.  Rather, we simply use it as a convenient way to quantify overlap between droplets.}. 

As shown in Fig.~\ref{fig:vs_as}, despite the rapid reduction of $\mathcal{C}$ with $a_s$, the simulated $f_{\rm{osc}}$ exhibits a rather constant behavior with a value always close to the purely irrotational predictions, $f_{\rm{irr}}$, for both a linear (1D) and hexagon state (2D). This observation indicates that (i) the degree of inter-droplet connection is not actually a major determinant of the angular oscillation frequency and (ii) that the system does not undergo RBR even for vanishingly small inter-droplet density connection. The latter conclusion is particularly evident for hexagon states, where the rigid-body prediction substantially departs from the irrotational one.
For the linear array, the elongated geometry means that the $f_{\rm{rig}}$ and $f_{\rm{irr}}$ differ only slightly; see supplemental materials for further discussion  \cite{supmat}.

\begin{figure*}[!htb]
\includegraphics[width=7 in, ]{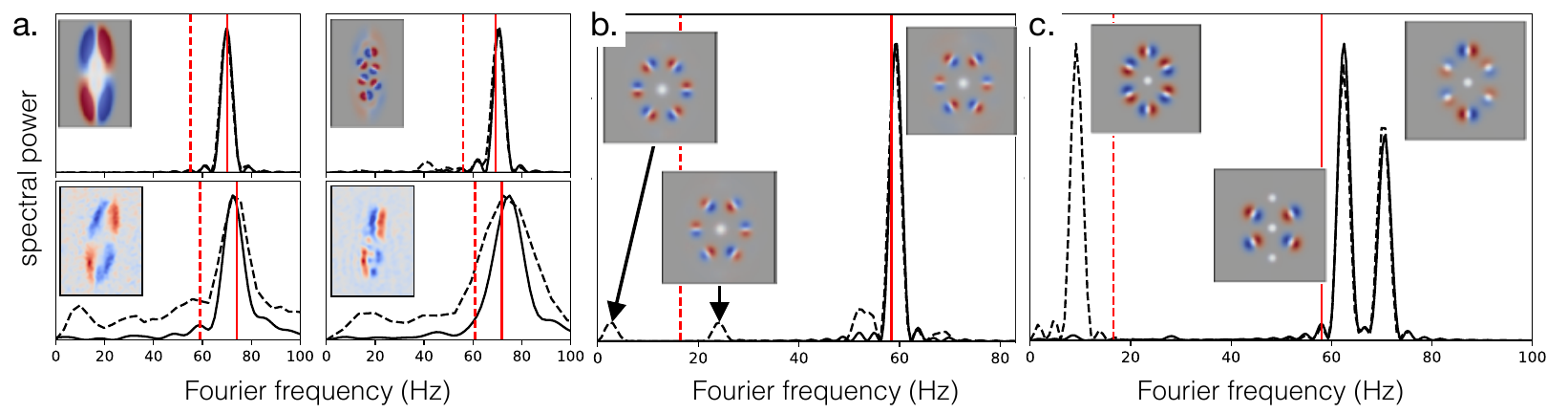}
\caption{Analysis of mode shapes and response due to angular excitation.  Solid lines are the power spectrum obtained from the rotational signal ($\theta$ in the experiment and $\langle xy \rangle$ in the simulation), and dashed lines are obtained from FTIA (see text, supplementary materials \cite{supmat} for description). Inset panels show the mode shapes for selected modes.  Red and blue indicate out-of-phase changes in density, overlaid onto the average density profile in the panels corresponding to simulation (gray to white). Solid and dashed vertical red lines represent $f_{\rm{irr}}$ and $f_{\rm{rig}}$, respectively. \textbf{a.} Responses in elongated traps from simulation (top) and experiment (bottom), for an unmodulated BEC (left) and a zigzag droplet state (right). Trap frequencies are $f_{x,y} = [31(1), 73(1), 128(1)]$~Hz, and $f_{x,y,z} = [32, 70, 122]$~Hz for the experiment and theory, respectively. \textbf{b.}  Response of supersolid hexagon state ($a_s = 92 a_0$). \textbf{c.} Response of droplet crystal hexagon state ($a_s = 85 a_0$). Note that the ground state has a different rotation for the two scattering lengths in this trap.  Trap frequencies are $f_{x,y,z} = [43, 53, 122]$~Hz for b,c.  }
\label{fig:profiles}
\end{figure*}

To better understand the non-rigid nature of the angular oscillations, we develop a method to extract the character of the system's response by analyzing our experimental and eGPE simulation dynamics in the frequency domain with respect to time, but in the position domain with respect to the spatial coordinates.
This technique, which we call ``Fourier transform image analysis" (FTIA) and detail in the supplemental materials \cite{supmat}, allows us to extract a power spectrum of density fluctuations driven by the angular excitation, as well as the spatial form of the density fluctuations at each frequency.  For comparison, we also calculate the spectral power of our rotational signal through a Fourier transform. For computational robustness, we use the fitted angle $\theta$ in the experimental case, and $\langle xy \rangle$ for the simulations.  To enhance our frequency resolution, we analyze simulations with longer durations than are accessible in the experiment (160 to 290~ms).

We apply the FTIA to both simulation and experimental images in Fig.~\ref{fig:profiles}a).  For a BEC the FTIA gives a dominant peak in both simulation and experiment, whose frequency and shape are consistent with a scissors mode oscillation at the frequency observed from the angular response.  For a zig-zag modulated state, we again predominantly observe a single peak in the FTIA spectrum at the frequency of the angular oscillation.  
In the simulation, we can see that the mode corresponds to the motion of the different droplets in a pattern reminiscent of IF in an unmodulated superfluid, and clearly distinct from RBR. In the experiment, the response of individual droplets is not visible, due to shot-to-shot fluctuations in the exact number and position of the droplets, but the overall structure is similar to the simulation.

For hexagonal supersolid (Fig.~\ref{fig:profiles}b) and isolated droplet (Fig.~\ref{fig:profiles}c) states, the FTIA reveals a clear multi-frequency response.  For the supersolid, we observe the excitation of modes near 3~Hz and 25~Hz that do not contribute strongly to $\langle xy \rangle$.  The droplet motion associated with the 3~Hz mode is approximately (but not exactly) shape-preserving, and the frequency is much lower than would be expected for a single-mode RBR response.  
For the isolated droplet array, we again observe a nearly-shape-preserving low-frequency response from FTIA, as well as a dominant angular response that is split into two frequencies, both above the scissors mode frequency $f_{\rm{irr}}$ expected for a superfluid with the same geometry. 
In the experiment, the combination of non-angular excitations associated with our method used to rotate the trap and relatively rapid damping of the oscillation prevent us from observing meaningful mode profiles for small $\beta$.

Importantly, the FTIA reveals that even in cases where we observe an apparently single-frequency response in typical rotational observables like $\theta$ or $\langle xy \rangle$ (as in Fig.~\ref{fig:profiles}a,b), the response of the system may in fact be multimode in nature, breaking the single-mode approximation used to analytically extract $f_{\rm{irr}}$ and $f_{\rm{irr}}$ \cite{pitaevskii2016bose, Roccuzzo:2020}.     In the case of a multi-frequency response, $f_{\rm{irr}}$ and $f_{\rm{irr}}$ instead provide an upper bound for the frequency of the lowest energy excitation -- an excitation that is difficult to see with experimentally accessible observables.  

Surprisingly, not only does the dominant angular response frequency fail to approach the rigid-body value in the isolated droplet regime, but it also stays near to the irrotational prediction.  A possible intuitive explanation for this observation is that the flow pattern of Fig.~\ref{fig:overview}a resembles that of a quadrupolar surface mode, and it is well-known that the frequency of such modes is predominantly determined by the trap parameters, rather than the details of the interparticle interactions \cite{pitaevskii2016bose}.  

In conclusion, measurements of angular oscillation frequencies offer a simple way to demonstrate superfluidity in certain conditions.  However, care must be taken when making and interpreting such measurements --- geometrical changes can mask the effects of changing superfluidity, and usual predictions to which one might compare rely on the assumption of a single-frequency response of the lowest energy rotational mode.  We find that a supersolid with 2D structure, which one might expect to be an ideal candidate for such measurements, can exhibit an apparently single-frequency response associated with a mode that is not the lowest in energy.  Further, this excitation frequency is typically very close to that of a purely superfluid system, even for systems where the effects of superfluidity are minimal. Therefore such measurements do not provide a robust indicator of superfluidity for modulated systems.  In the future, it may be possible to extract information about superfluidity using a modified excitation scheme to preferentially excite the lower energy modes, and a more comprehensive analysis scheme suitable for multi-frequency response \cite{pc}.  However, such techniques would require detailed knowledge of the exact excitation applied, and measurement of response amplitudes, both of which are considerably more challenging in an experiment than measuring the frequency of an oscillation.  

Finally, we note that even in the case of single-frequency response, where the frequency of angular oscillations has a direct connection to the moment of inertia of the system, making a clear connection between the moment of inertia and quantities like a superfluid fraction can be problematic.  Past works have predicted that a system which is partially superfluid should have a moment of inertia in between the RBR and IF predictions, linearly interpolated according to a superfluid fraction \cite{Leggett:1970, tanzi2021evidence}.  While this interpretation may be valid for systems featuring a rigid crystaline structure and a uniform distribution of crystaline and superfluid components, as in \cite{Leggett:1970}, it is not necessarily valid for our small dipolar supersolids, which in addition to coupled superfluid-crystaline excitations, feature a nonuniform degree of modulation across the system.

\textbf{Acknowledgments:}
We thank Sandro Stringari and Alessio Recati for useful discussions. We acknowledge R.M.W. van Bijnen for developing the code for our eGPE ground-state simulations.
The experimental team is financially supported through an ERC Consolidator Grant (RARE, No.\,681432), an NFRI grant (MIRARE, No.\,\"OAW0600) of the Austrian Academy of Science, the QuantERA grant MAQS by the Austrian Science Fund FWF No\,I4391-N. L.S and F.F. acknowledge the DFG/FWF via FOR 2247/PI2790. 
L.S.~ thanks the funding by the Deutsche Forschungsgemeinschaft (DFG, German Research Foundation) under Germany’s Excellence Strategy – EXC-2123 QuantumFrontiers – 390837967. M.A.N.~has received funding as an ESQ Postdoctoral Fellow from the European Union’s Horizon 2020 research and innovation programme under the Marie Skłodowska‐Curie grant agreement No.~801110 and the Austrian Federal Ministry of Education, Science and Research (BMBWF). M.J.M.~acknowledges support through an ESQ Discovery Grant by the Austrian Academy of Sciences.  
We also acknowledge the Innsbruck Laser Core Facility, financed by the Austrian Federal Ministry of Science, Research and Economy. Part of the computational results presented have been achieved using the HPC infrastructure LEO of the University of Innsbruck.

\bibliography{references}
\end{document}


\renewcommand\thefigure{\thesection S\arabic{figure}}   
\setcounter{figure}{0}   
\title{Supplemental materials for: Can angular oscillations probe superfluidity in dipolar supersolids?}

 \author{Matthew A. Norcia}
 \affiliation{
     Institut f\"{u}r Quantenoptik und Quanteninformation, \"Osterreichische Akademie der Wissenschaften, Innsbruck, Austria
 }

\author{Elena Poli}
 \affiliation{
     Institut f\"{u}r Experimentalphysik, Universit\"{a}t Innsbruck, Austria
 }
 
 \author{Claudia Politi}
 \affiliation{
     Institut f\"{u}r Quantenoptik und Quanteninformation, \"Osterreichische Akademie der Wissenschaften, Innsbruck, Austria
 }
 \affiliation{
     Institut f\"{u}r Experimentalphysik, Universit\"{a}t Innsbruck, Austria
 }

 \author{Lauritz Klaus}
 \affiliation{
     Institut f\"{u}r Quantenoptik und Quanteninformation, \"Osterreichische Akademie der Wissenschaften, Innsbruck, Austria
 }
 \affiliation{
     Institut f\"{u}r Experimentalphysik, Universit\"{a}t Innsbruck, Austria
 }

\author{Thomas Bland}
 \affiliation{
     Institut f\"{u}r Quantenoptik und Quanteninformation, \"Osterreichische Akademie der Wissenschaften, Innsbruck, Austria
 }
 \affiliation{
     Institut f\"{u}r Experimentalphysik, Universit\"{a}t Innsbruck, Austria
 }
 
 \author{Manfred J. Mark}
 \affiliation{
     Institut f\"{u}r Quantenoptik und Quanteninformation, \"Osterreichische Akademie der Wissenschaften, Innsbruck, Austria
 }
 \affiliation{
     Institut f\"{u}r Experimentalphysik, Universit\"{a}t Innsbruck, Austria
 }

 \author{Luis Santos}
 \affiliation{
     Institut f\"{u}r Theoretische Physik, Leibniz Universit\"{a}t Hannover, Germany
 }
 
\author{Russell N. Bisset}
 \affiliation{
     Institut f\"{u}r Experimentalphysik, Universit\"{a}t Innsbruck, Austria
 }
 
 \author{Francesca Ferlaino}
 \thanks{Correspondence should be addressed to \mbox{\url{Francesca.Ferlaino@uibk.ac.at}}}
 \affiliation{
     Institut f\"{u}r Quantenoptik und Quanteninformation, \"Osterreichische Akademie der Wissenschaften, Innsbruck, Austria
 }
 \affiliation{
     Institut f\"{u}r Experimentalphysik, Universit\"{a}t Innsbruck, Austria
 }

\maketitle

\section{Excitation protocol}
In both experiment and simulation, we excite the atoms by suddenly rotating the trap, holding for 6~ms, then returning it to its initial orientation.  This was important in the experiment, as the trap frequencies generally change slightly as the trap is rotated, and we want to observe the evolution of a state that is equilibrated to the trap prior to the rotation.  
To explore whether the exact excitation protocol influences our results, we performed additional simulations where the trap angle was rotated and held in the new orientation, but not rotated back.  We find that the same modes are excited in this case, and the frequency of their responses are the same.  For some parameters the relative contributions of the modes to the spectrum of $\langle xy \rangle$ can differ between the two protocols, but for the parameters we explore the frequency of the peak response remains unchanged.  In particular, for the droplet crystal hexagon shown in Fig.~4c of the main text, the contribution of the low-frequency mode to the $\langle xy \rangle$ power spectrum becomes appreciable, though is still smaller than the contribution of the modes near 60~Hz.  

We have also performed excitation in the simulation by directly imprinting a small phase variation $\alpha x y$ onto the ground-state wavefunction.  This protocol produces qualitatively similar results to those described above.  Again, the same modes are excited and respond with the same frequencies, though sometimes with different amplitudes.  The dominant mode excited is the same as the rotate-and-return protocol for all cases investigated.  

\section{Extracting angular power spectrum}
Several methods can be used to extract the angular response of our system.  For the experiment, we perform a two-dimensional Gaussian fit to the in-trap image, and record the angle of the major and minor axes as a function of time.  For the simulation, we report the angular response obtained using one of two observables.  For direct comparison to the experiment, we use the state angle extracted from a 2D Gaussian fit, as in the experiment. For more detailed spectral analysis, we use the quantity $\langle xy \rangle$, as this is expected to have a strong response to a rapid rotation of the trap and we find it to be numerically more robust.  We have confirmed that these and other similar observables, such as the directions of maximal and minimal variance, provide consistent results (up to overall normalization).  In some cases, the Fourier spectrum of $\langle \hat{L}_z \rangle$ (though not experimentally accessible) shows different relative response amplitudes between modes compared to $\langle xy \rangle$, particularly for those modes at low frequencies.

\section{Fourier transform image analysis}

\begin{figure}[!htb]
\includegraphics[width=3.38 in, ]{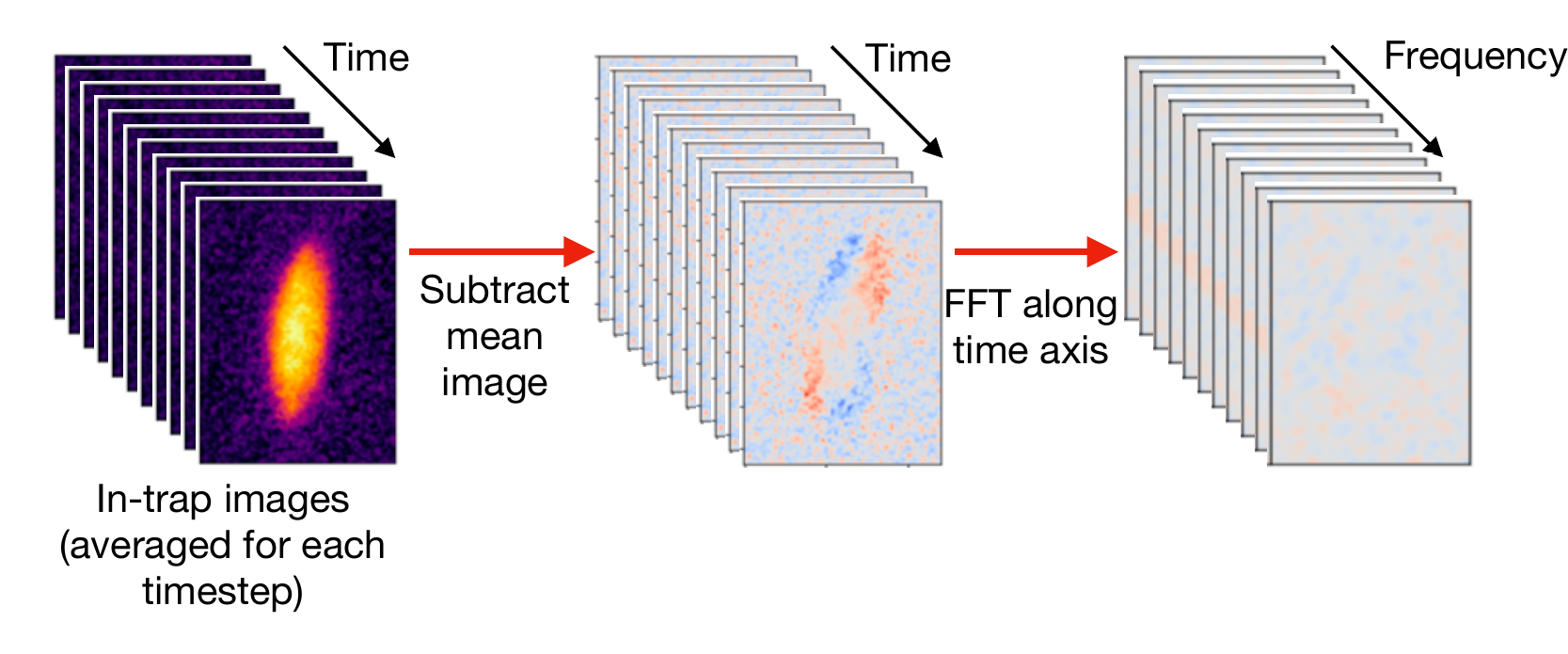}
\caption{Procedure for Fourier transform image analysis (FTIA).  See text for description.     }
\label{fig:ftia}
\end{figure}

The goal of our Fourier transform image analysis (FTIA) protocol is to visualize the density response of our atomic system in real-space with respect to position, but in frequency space with respect to time.  This provides a simple way to extract the spatial profile of excitation modes.  The process is illustrated in Fig.~\ref{fig:ftia}.  
To perform the FTIA, we assemble images of projected density profiles corresponding to single time-steps (directly from the simulation, or averaged over several in-trap images from the experiment), then subtract the average (over all time-steps) image from each.  We then Fourier transform the results along the time axis.  The output is then a sequence of real-space images, showing the fluctuation pattern at a given frequency. Because each pixel is now represented by a complex number (encoding the amplitude and phase of the density variations at that location), we plot with respect to the global phase for each frequency that shows maximum variation, thus plotting the in-phase quadrature of the oscillation. 

In order to obtain a power spectrum (useful for locating the frequencies of excited modes), we compute the sum of the absolute square of the fluctuations for each frequency.  
This power spectrum can be used to identify the frequency and spatial character of modes, but is not expressed in physically meaningful units, and so should not necessarily be used to compare the strength of different mode responses. 

We note that there are some similarities between the FTIA method and principle component analysis (PCA) \cite{jolliffe2002generalizations}.  Both provide a model-free way of extracting the form of excitations present in a system.  PCA does so by finding correlated patterns of fluctuations within a set of images, with no prior information about the time-sequence of the images.  This makes it well-suited to revealing modes that are excited incoherently, for example by thermal or quantum noise.  In contrast, our FTIA method explicitly incorporates the time-domain information associated with the images.  This makes it well-suited to extracting modes that are coherently excited (FTIA, as we apply it, would not work for incoherently excited modes).  In practice, we find that the FTIA is more robust than PCA at extracting fluctuation patterns that each exhibit a single-frequency response.  While PCA often returns components whose weights vary with multiple frequencies (indicating that they actually correspond to a linear combination of eigenmodes), FTIA by construction returns a fluctuation pattern associated with a single frequency.  
We find that this feature makes it more robust for identifying eigenmodes of a system subject to a coherent drive.

\section{spectra/table for all params}
\begin{figure*}[!htb]
\includegraphics[width=6.75 in, ]{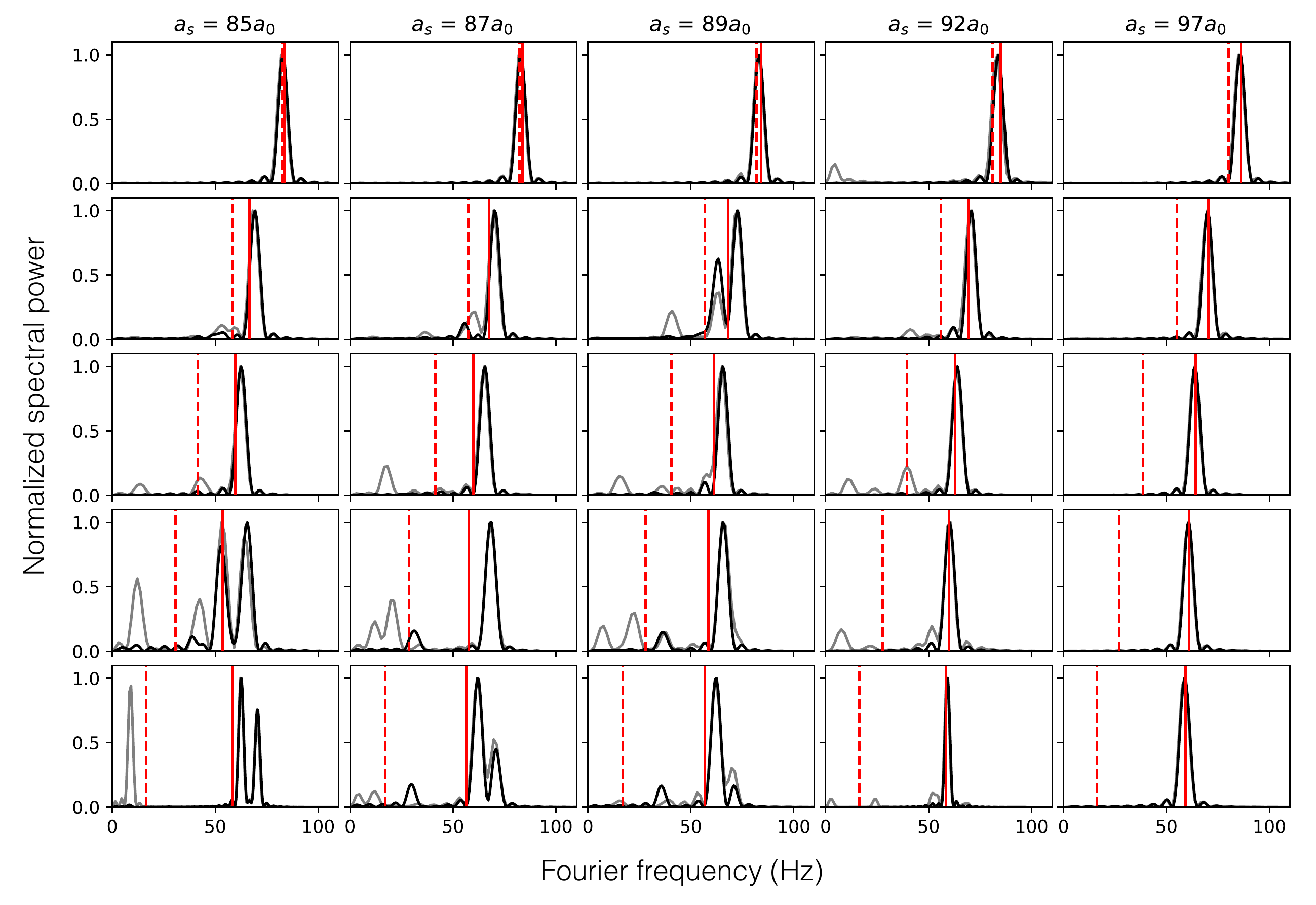}
\caption{Response sepctra extracted from simulations for different trap parameters (rows) and scattering lengths (columns).  Upper rows correspond to more elongated traps, while lower rows correspond to more round ones.  From top to bottom, $(f_x, f_y) = [(26, 87), (32, 70), (37,62), (40,57), (43,53)]$~Hz.  $f_z = 122$~Hz for all cases.  Red vertical dashed and solid lines correspond to the rigid-body rotation and irrotational flow predictions, respectively.  Gray traces are power spectra extracted from FTIA, while black traces are from $\langle xy \rangle$.  In all cases, $a_s = 97a_0$ corresponds to an unmodulated BEC, while lower scattering lengths correspond to modulated states, with the overlap between droplets decreasing with scattering length.  $(f_x, f_y) = (26, 87)$ is a linear droplet chain for all scattering lengths that produce a modulated state.  All other modulated states have transverse structure, increasing in prevalence as the trap becomes more round.  }
\label{fig:allsets}
\end{figure*}

Excitation power spectra from simulation for a range of traps and scattering lengths used in the main manuscript can be found in Fig.~\ref{fig:allsets}.  

\section{predictions for rotational mode frequencies}
The rotational response of a gas can be calculated using hydrodynamic equations \cite{guery1999scissors} or a  sum-rule approach \cite{pitaevskii2016bose, Roccuzzo:2020}.  From the sum-rule approach, an expression can be derived for the rotational oscillation frequency, under the assumption that the response is single-frequency:
\begin{equation}
\omega^2 = \frac{m \langle y^2 - x^2 \rangle (\omega_x^2 - \omega_y^2)}{\Theta}
\label{eq:1}
\end{equation}
Here, $\Theta$ is the moment of inertia associated with steady-state rotation.  

The numerator of Eq.~\ref{eq:1} can be interpreted as a rotational spring constant: $k_\tau = -\tau/\theta$, where $\tau$ is the torque exerted on a state whose major and minor axes $y$ and $x$ are rotated relative to the their equilibrium position in the trap by an angle $\theta$.  To see this, consider a mass element $m$ at position $(x, y)$ in a trapping potential $V = (m\omega_x x^2 + m\omega_y y^2)/2$, which exerts a torque  $\tau = x F_y - y F_x  = xym(\omega_x^2 - \omega_y^2)$.  We can then calculate $k_\tau = - \partial \tau/\partial \theta = -m(y\partial x/\partial \theta + x\partial y/\partial \theta)(\omega_x^2 - \omega_y^2) = m (y^2 - x^2)(\omega_x^2 - \omega_y^2)$. Summing over mass elements provides the numerator of Eq.~\ref{eq:1}.  This highlights that the numerator of this expression is purely geometrical, independent of whether the state is superfluid or classical.  
In the case of multi-frequency response, Eq.~\ref{eq:1} (as defined by the sum rule) becomes an inequality, defining the upper bound for the lowest frequency angular excitation in the system \cite{pitaevskii2016bose}.  


\section{beta versus scattering length for 1D and 2D}
\begin{figure}[!htb]
\includegraphics[width=3.38 in, ]{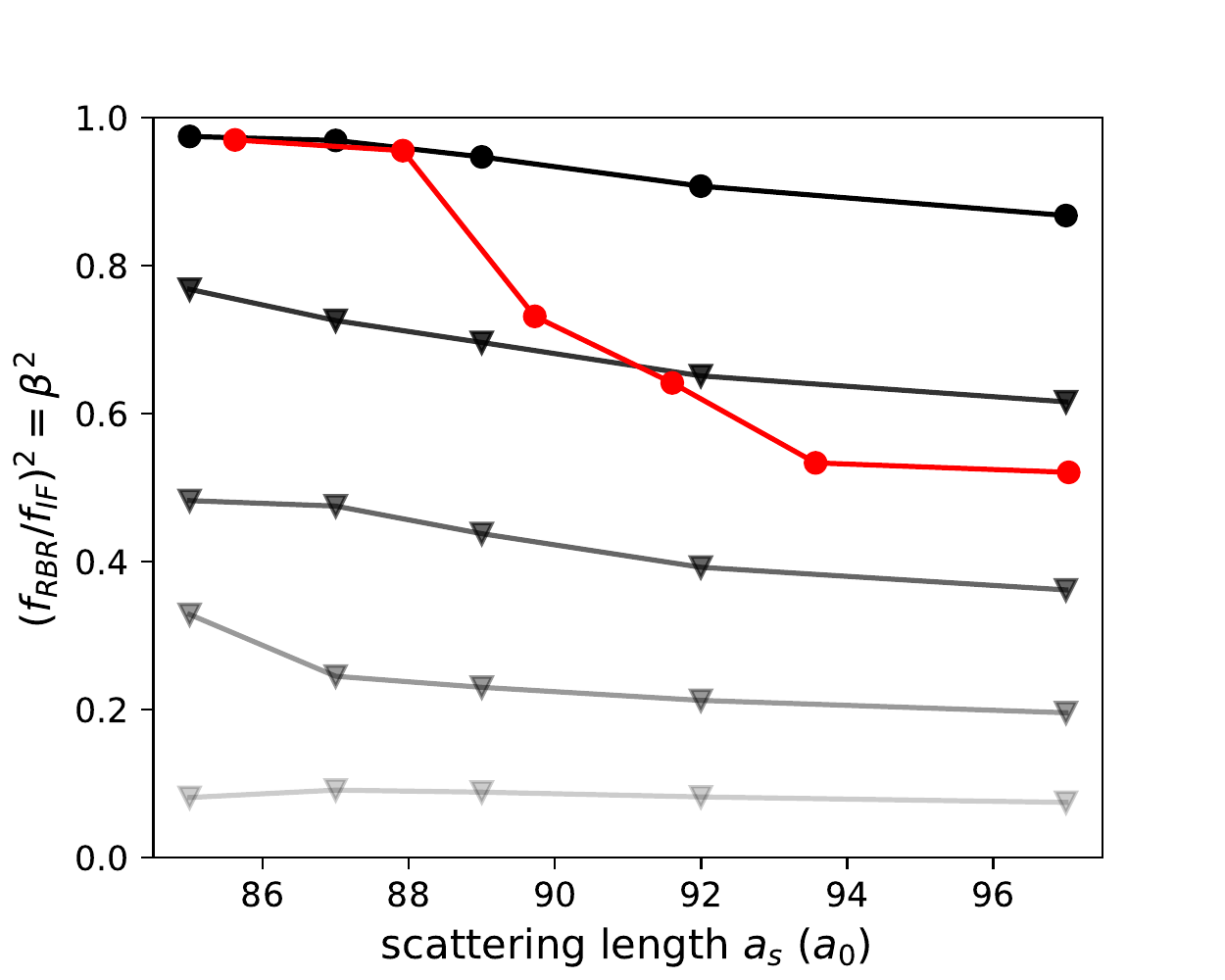}
\caption{Difference in moment of inertia between IF and RBR, $\beta^2$ for different traps and scattering lengths.  Traps and atom numbers that correspond to linear arrays for modulated states are indicated by round markers, while those that correspond to 2D modulated arrays are indicated by triangles.
All points except those at $a_s = 97 a_0$ are modulated, interdroplet connection decreasing with scattering length.  The states explored in Fig.~\ref{fig:allsets} are shown in grey-scale, with lighter saturation corresponding to rounder traps, while the conditions similar to those of Refs.~\cite{Roccuzzo:2020, tanzi2021evidence} are shown in red.    }
\label{fig:betas}
\end{figure}

In Fig.~\ref{fig:betas}, we show the change in the anisotropy of the atomic state in response to a change in scattering length for a variety of traps, featuring both linear and 2D array modulated configurations.  Here, we consider the quantity $\beta^2 = (\langle x^2 - y^2 \rangle / \langle x^2 + y^2\rangle)^2$, as this quantity gives the expected change in moment of inertia between irrotational flow (IF) and rigid-body rotation (RBR).  As $\beta^2$ approached unity, the difference between the two vanishes, so such states can exhibit minimal sensitivity to superfluidity.   

States in more elongated traps generally lead to more elongated states with values of $\beta^2$ closer to one than their rounder counterparts.  However, even in relatively round traps, such as those of Refs.~\cite{Roccuzzo:2020, tanzi2021evidence}, low atom numbers can lead to the formation of linear arrays, which are highly elongated.  In these cases, the  sensitivity of the state to superfluidity is dramatically reduced upon entering the modulated regime.  In contrast, combinations of trap parameters and atom number that lead to a 2D modulated state typically maintain values of $\beta^2$ substantially different from one even in the low scattering length, independent droplet regime.

\section{linear case}
\begin{figure}[!htb]
\includegraphics[width=3.38 in, ]{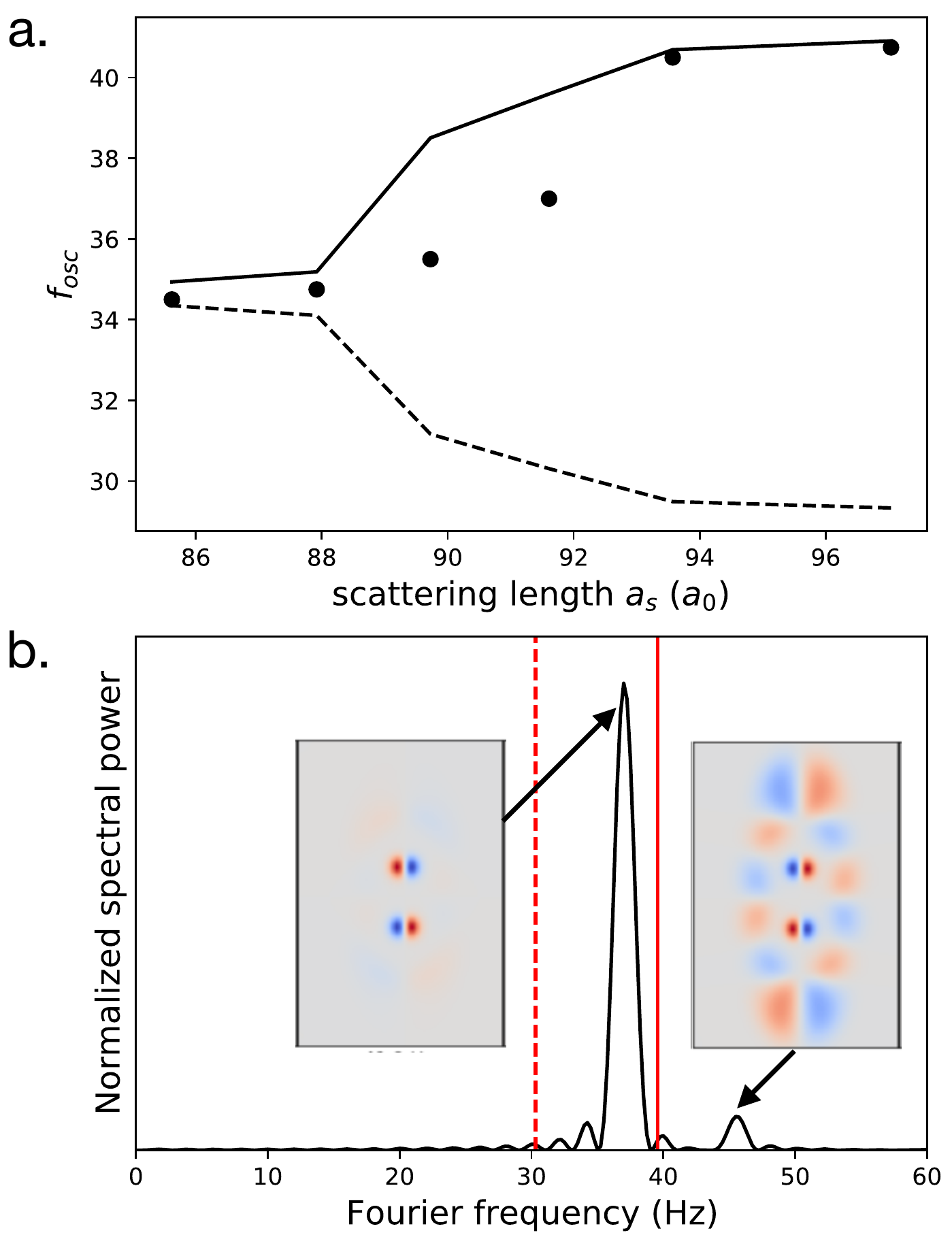}
\caption{Analysis of linear two-droplet arrays of \cite{Roccuzzo:2020, tanzi2021evidence}.  \textbf{a.} Dominant angular oscillation frequency (markers) extracted from simulations versus scattering length, through transition from BEC (right) to isolated droplets (left).  The irrotational and rigid-body predictions $f_{\rm{irr}}$ and $f_{\rm{rig}}$ are shown as solid and dashed lines, respectively.   \textbf{b.} The Fourier spectrum of $\langle xy \rangle$ for the point near $a_s = 92a_0$ exhibits a response with dominant and sub-dominant mode contributions.  The fluctuation profiles associated with these two frequencies are shown in the insets.  Solid and dashed vertical red lines represent $f_{\rm{irr}}$ and $f_{\rm{rig}}$, respectively.  }
\label{fig:linear}
\end{figure}

In Fig.~\ref{fig:linear}, we explore the parameters of refs~\cite{Roccuzzo:2020, tanzi2021evidence}, where a change in scattering length induces a transition from an unmodulated BEC to a linear array of two droplets.  This transition is accompanied by a dramatic change in the aspect ratio of the atomic state, as evident in the near convergence of the predictions for rigid body and irrotational flow ($f_{\rm{rig}}$ and $f_{\rm{irr}}$) at lower scattering lengths, corresponding to the droplet state.  

We see that the dominant frequency of angular response is between $f_{\rm{rig}}$ and $f_{\rm{irr}}$, indicating a change in the level of superfluidity in the system.  We find that the angular response in the supersolid regime ($a_s$ = 90 or 92 $a_0$) has two clear frequency components, though in this case the dominant frequency observed matches the prediction from the sum rule (with moment of inertia calculated under static rotation).  Because of the geometry of the system, rotation does not lead to a significant transfer of mass between the two droplets.  Thus, we attribute the change in superfluidity to the low-density halo that surrounds the droplets, rather than the inter-droplet connection itself.

\bibliography{references}